\documentclass[submission,copyright,creativecommons]{eptcs}
\providecommand{\event}{GASCOm 2024} 

\usepackage{iftex}

\ifpdf
  \usepackage{underscore}         
  \usepackage[T1]{fontenc}        
\else
  \usepackage{breakurl}           
\fi

\title{Restricted Permutations Enumerated by Inversions}
\author{Atli Fannar Franklín
\institute{Department of Mathematics\\
University of Iceland\\
Reykjavík, Iceland}
\email{aff6@hi.is}
\and
Anders Claesson
\institute{Department of Mathematics\\
University of Iceland\\
Reykjavík, Iceland}
\email{akc@hi.is}
\and
Christian Bean
\institute{School of Computer Science
and Mathematics\\
Keele University\\
Keele, United Kingdon}
\email{\quad c.n.bean@keele.ac.uk}
\and
Henning Úlfarsson
\institute{Department of Computer Science\\
Reykjavík University\\
Reykjavík, Iceland}
\email{\quad henningu@ru.is}
\and
Jay Pantone
\institute{Department of Mathematical
and Statistical Sciences\\
Marquette University\\
Milwaukee, WI, USA}
\email{\quad jay.pantone@marquette.edu}
}
\def\titlerunning{Restricted Permutatations Enumerated by Inversions}
\def\authorrunning{A. F. Franklín, A. Claesson, C. Bean. H. Úlfarsson, J. Pantone}

\newtheorem{theorem}{Theorem}
\newtheorem{lemma}[theorem]{Lemma}
\usepackage{tikz}
\usetikzlibrary{matrix,arrows,patterns,calc,positioning}
\usepackage{amssymb,amsmath}
\newcommand\p[1]{\left(#1\right)}

\begin{document}
\maketitle

\begin{abstract}
Permutations are usually enumerated by size, but new results can be found by enumerating them by inversions instead, in which case one
  must restrict one's attention to indecomposable permutations. In the style of the seminal paper by Simion and Schmidt \cite{seminal}, we investigate all
  combinations of permutation patterns of length at most $3$.
\end{abstract}

\section{Introduction}

To enumerate permutations by their number of inversions we need the set of permutations with a given number of inversions to be finite. This is generally not the case, since in many cases one can add a new maximal element to the end of the permutation to get a new one with equally many inversions. To get around this we introduce the notion of decomposability. We say that a permutation $\pi = \pi_1 \dots \pi_n$ is decomposable if there exists an index $i < n$ such that $\pi_1 \dots \pi_i$ is a permutation of the elements $1, \dots, i$. If we let $\rho = \pi_1 \dots \pi_i$ and $\tau = (\pi_{i+1}-i) \dots (\pi_n-i)$ we denote this $\pi = \rho \oplus \tau$. In the same vein, permutations that are not decomposable are called indecomposable. With this definition every permutation can be factored into a set of indecomposable factors, and we call these factors its components.

An inversion in a permutation $\pi = \pi_1 \dots \pi_n$ is a pair of indices $(i, j)$ such that $i < j$ and $\pi_i > \pi_j$. The number of inversions in a permutation $\pi$ will be denoted $\operatorname{inv}(\pi)$. The following lemma highlights a relation between these two concepts.

\begin{lemma}\label{invcomp}
Let $\pi$ be a permutation on $n$ elements and $c$ components. Then $\operatorname{inv}(\pi) \geq n - c$.
\end{lemma}

Crucially this means that if $c = 1$ then $\operatorname{inv}(\pi) \geq n - 1$. Thus the set of indecomposable permutations with $k$ inversions is finite since it is contained in the set of all permutations on $k + 1$ elements or fewer. With this in mind, we define $I_k$ to be the set of all indecomposable permutations with exactly $k$ inversions.

Next we recall some notions related to patterns of permutations. We say two sequences $a_1, \dots, a_m$ and $b_1, \dots, b_m$ are order-isomorphic if $a_i < a_j$ holds if and only if $b_i < b_j$. For permutations $\pi = \pi_1 \dots \pi_n$, $\tau = \tau_1 \dots \tau_m$ we say that $\pi$ contains the pattern $\tau$ if there exists a set of indices $i_1, \dots, i_m$ such that $\pi_{i_1} \dots \pi_{i_m}$ is order-isomorphic to $\tau_1 \dots \tau_m$. We call such a set of indices an \emph{occurrence} of the pattern $\tau$. If $\pi$ has no occurrence of $\tau$ we will say that $\pi$ \emph{avoids} the pattern $\tau$. We will denote the subset of $I_k$ containing the permutations avoiding $\tau$ by $I_k(\tau)$. Similarly we will denote the subset of $I_k$ containing permutations avoiding several patterns $\tau_1, \tau_2, \dots, \tau_r$ by $I_k(\tau_1, \tau_2, \dots, \tau_r)$.

\section{Single patterns}

As examples we have that $I_k(1)$ and $I_k(21)$ have no non-empty elements and $|I_k(12)|$ is the characteristic function of the triangular numbers, listed in the OEIS as \texttt{A010054}. The remaining single patterns of length $3$ are then $123, 132, 213, 231, 312$ and $321$. We consider the reverse complement of a permutation: written out explicitly, this is $\pi_1 \dots \pi_n \mapsto (n + 1 - \pi_n)(n + 1 - \pi_{n-1})\dots(n + 1 - \pi_1)$. Consider an inversion on $i < j$ in $\pi$. The values $\pi_i > \pi_j$ get mapped to $n + 1 - \pi_i, n + 1 - \pi_j$ at indices $n + 1 - i, n + 1 - j$ in the image. Thus the size of the elements is inverted, but so is their order. Therefore the number of inversions remains constant under reverse complement. We also see that $\pi$ is decomposable if and only if its reverse complement is, so the reverse complement is an involution on $I_k$ for every $k$. It maps the pattern $132$ to the pattern $213$, so from this we see that $|I_k(213)| = |I_k(132)|$, hence we only have to consider one of these patterns. This is similar to what happens when counting pattern avoiding permutations classically, however we must compose reversion and complement for the argument to work. Normally this line of reasoning shows that the number of $132$ and $231$ avoiding permutations are the same through reversion, but we will see that this is no longer the case when enumerating by inversions. A similar argument can be made to show that mapping to the inverse permutation preserves inversions and indecomposability as well, and in fact the symmetries are generated by these two maps.

This means we only have to consider the patterns $123, 132, 231$ and $321$. We will start with $132$. To do this, we make use of a well known bijection on permutations. The inversion table of a permutation $\pi = \pi_1 \dots \pi_n$ is the sequence $b_1b_2 \dots b_n$, where $b_i$ is the number of values after $\pi_i$ in $\pi$ that are smaller than $\pi_i$. The image of this bijection is given by the set of all sequences such that the first value is $\leq n - 1$, the next $\leq n - 2$ and so on. We call such sequences subdiagonal. Furthermore, elements in a subdiagonal sequence that are equal to their maximum possible value are called diagonal elements.

\begin{theorem}\label{partitions}
$|I_k(132)|$ counts partitions on $k$ elements, listed as \texttt{A000041} on the OEIS.
\end{theorem}

To work with $231$-avoiding permutations we define the \emph{skew-sum} of two permutations $\pi_1 \dots \pi_n$ and $\tau_1 \dots \tau_m$, as the permutation on $n + m$ elements given by $\pi \ominus \tau = (\pi_1 + m) \dots (\pi_n + m) \tau_1 \dots \tau_m$. We will say that $\pi$ is skew-decomposable if it can be written as $\pi = \tau_1 \ominus \tau_2$ for non-empty $\tau_1, \tau_2$, and skew-indecomposable otherwise.

\begin{theorem}
$|I_k(231)|$ counts fountains on $k$ coins. A fountain of coins is an arrangement of coins in rows such that the bottom row is full (that is, there are no ``holes''), and such that each coin in a higher row rests on two coins in the row below. This is listed as \texttt{A005169} on the OEIS.
\end{theorem}

Enumerating $321$-avoiding permutations by size and inversions (allowing decomposable permutations) has been investigated in \cite{321}. A generating function is derived, but as we need a bijective map from $I_k(321)$ to our target set for what comes later, we will have to give a different proof.

\begin{theorem}\label{ik321}
$|I_k(321)|$ counts parallelogram polyominoes with $k$ cells, listed as \texttt{A006958} on the OEIS.
\end{theorem}

This alternative view on parallelogram polyominoes opens up a formula for efficiently computing new terms of the series, taking $\mathcal{O}(k^2)$ time and space to compute $|I_k(321)|$.

\begin{theorem}\label{321rec}
$|I_k(321)| = a_{k, 1}$ where 
\[a_{n,m} = \begin{cases}
1 \text{ if } n = 0 \\
\sum_{i = 1}^n a_{n - i, i} \text{ if } m = 1 \\
a_{n, m - 1} + \sum_{i = m}^n a_{n - i, i} \text{ otherwise} \\
\end{cases}\]
\end{theorem}

Not only can this view help with computing new terms, but it also gives rise to new bijective correspondences. Consider fountains of coins where we only count coins in even rows. We can still place coins as we like, but when tallying the number we only count those in the bottom row, those 2 rows up, 4 rows up and so on. Such fountains with $n$ counted coins will be called even fountains of size $n$. In \cite{fountain} the problem of mapping parallelogram polyominoes to such fountains is tackled. It is shown that there are equally many by an algebraic argument, but it is left as an open question at the end whether there is any bijective proof. Using $I_k(321)$, a bijective proof can be found. Since Theorem \ref{ik321} is proved by mapping parallelogram polyominoes bijectively to $I_k(321)$, it suffices to map $I_k(321)$ bijectively to even fountains of size $k$.

\begin{theorem}
$I_k(321)$ maps bijectively to even fountains of size $k$.
\end{theorem}

The proof of Theorem \ref{321rec} is bijective, so it suffices to map the even fountains to the sequences described there. We will describe a map taking even fountains to such sequences.

Write out the fountain in the usual manner (see picture), calling coins in even rows red and the ones in odd rows black for convenience. For each coin in the bottom row, going from left to right, we do the following procedure repeatedly:

\begin{itemize}
\item If we are on a red coin, we remove it and move to the coin above and to the right. If there is no such coin we stop.
\item If we are on a black coin, we remove it and move to the coin above and to the right. If there is no such coin we move to the coin below and to the right instead.
\end{itemize}

Once done with a coin in the bottom row, we write down the number of red coins removed during this procedure. This produces a sequence of numbers, which we then append a single zero to. We claim this produces a sequence of the desired form.

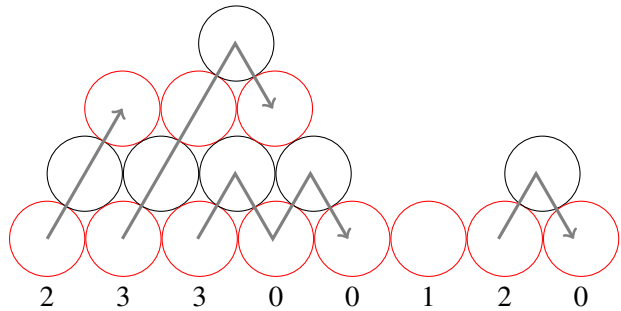
\begin{center}
\begin{tikzpicture}
\node[circle,draw,minimum size=1cm, red](r1c1) {};
\node[circle,draw,minimum size=1cm](r2c1) at (60:1cm) {};
\node[circle,draw,minimum size=1cm, red](r3c1) at (60:2cm) {};
\node[circle,minimum size=1cm](r4c1) at (60:3cm) {};

\node[circle,draw,minimum size=1cm, right=0mm of r1c1, red](r1c2) {};
\node[circle,draw,minimum size=1cm, right=0mm of r2c1](r2c2) {};
\node[circle,draw,minimum size=1cm, right=0mm of r3c1, red](r3c2) {};
\node[circle,draw,minimum size=1cm, right=0mm of r4c1](r4c2) {};

\node[circle,draw,minimum size=1cm, right=0mm of r1c2, red](r1c3) {};
\node[circle,draw,minimum size=1cm, right=0mm of r2c2](r2c3) {};
\node[circle,draw,minimum size=1cm, right=0mm of r3c2, red](r3c3) {};

\node[circle,draw,minimum size=1cm, right=0mm of r1c3, red](r1c4) {};
\node[circle,draw,minimum size=1cm, right=0mm of r2c3](r2c4) {};

\node[circle,draw,minimum size=1cm, right=0mm of r1c4, red](r1c5) {};
\node[circle,minimum size=1cm, right=0mm of r2c4](r2c5) {};

\node[circle,draw,minimum size=1cm, right=0mm of r1c5, red](r1c6) {};
\node[circle,minimum size=1cm, right=0mm of r2c5](r2c6) {};

\node[circle,draw,minimum size=1cm, right=0mm of r1c6, red](r1c7) {};
\node[circle,draw,minimum size=1cm, right=0mm of r2c6](r2c7) {};

\node[circle,draw,minimum size=1cm, right=0mm of r1c7, red](r1c8) {};

\draw[->, very thick, gray] (0, 0) -- ++(60:1cm) -- ++(60:1cm);
\draw[->, very thick, gray] (1, 0) -- ++(60:1cm) -- ++(60:1cm) -- ++(60:1cm) -- ++(-60:1cm);
\draw[->, very thick, gray] (2, 0) -- ++(60:1cm) -- ++(-60:1cm) -- ++(60:1cm) -- ++(-60:1cm);
\draw[->, very thick, gray] (6, 0) -- ++(60:1cm) -- ++(-60:1cm);

\node[below=0mm of r1c1] (t1) {$2$};
\node[below=0mm of r1c2] (t2) {$3$};
\node[below=0mm of r1c3] (t3) {$3$};
\node[below=0mm of r1c4] (t4) {$0$};
\node[below=0mm of r1c5] (t5) {$0$};
\node[below=0mm of r1c6] (t6) {$1$};
\node[below=0mm of r1c7] (t7) {$2$};
\node[below=0mm of r1c8] (t8) {$0$};
\end{tikzpicture}
\end{center}

This leaves us only with $I_k(123)$, the only single pattern giving rise to a sequence not in the OEIS.

\begin{theorem}
$|I_k(123)|$ counts indecomposable subdiagonal sequences where non-diagonal elements are in decreasing order. Let
\[c_{n,m,k} = 
\begin{cases} 
0 \text{ if } n < 0 \text{ or } k < 0 \\
1 \text{ if } n = k = 0 \\
c_{n - 1, m, l - n + 1} + \sum_{i = 0}^{\min(n - 2, m - 1)} c_{n - 1, i, k - i} \text{ otherwise }
\end{cases}\]
Then the total number of permutations with $k$ inversions (including decomposable ones) avoiding $123$ can be calculated as $\sum_{n = 0}^{k+1} c_{n, n, k}$.

To obtain the number of such permutations that are indecomposable, subtract the $k$-th coefficient of $\p{\sum_{i \geq 0} x^{i(i+1)/2}}^2$.
\end{theorem}

\section{Several patterns}

We now investigate permutations avoiding several patterns. Some groups of patterns are restrictive enough to make all supersets of those patterns trivially determined. For example $|I_k(123, 321)|$ quickly decays to zero by the Erdös-Szekeres theorem, making all supersets of $123, 321$ easy to determine. We have two more such pairs of patterns.

\begin{theorem}
$|I_k(231, 321)| = 1$ and the unique permutation with $k$ inversions is $k 1 2 \dots (k - 1)$.
\end{theorem}

\begin{theorem}
$|I_k(231, 312)| = |I_k(12)|$.
\end{theorem}

By utilizing the symmetries we have, the only pairs of patterns left to investigate are $123, 231$ and pairs containing $132$.

\begin{theorem}
$|I_k(123, 231)|$ counts fountains of $k$ coins where the missing coins with respect to a full triangular fountain form a rectangle (removing no coins counts as a rectangle). The generating function is given by $\sum_{i \geq 1} x^{\binom{i}{2}} + \sum_{i \geq 1} \sum_{j \geq 1} \sum_{\ell = 0}^{\min(i, j) - 1} x^{\binom{i + 1}{2} + \binom{j + 1}{2} - \binom{\ell + 1}{2}}$.
\end{theorem}

This leaves us with four pairs, pairing $132$ with any of the patterns $123, 213, 231$ and $321$. We now tackle them in that order.

\begin{theorem}
$|I_k(132, 123)|$ enumerates the Pascal triangle with the first column removed, which is listed as \texttt{A135278} on the OEIS. It has generating function $\sum_{n \geq 0} x^{n(n+3)/2} \p{(x + 1)^{n+2} - x^{n+2}}$.
\end{theorem}

Our next result involves a kind of partition called a Gorenstein partition. Gorenstein partitions are partitions whose maximal chains are all of the same size when regarded as order ideals of $\{1, 2, \dots\} \times \{1, 2, \dots\}$. This definition is rather unwieldy for our purposes, so we first translate this condition.

\begin{lemma}
\label{gorenlemma}
A partition $\rho$ is Gorenstein if and only if $\rho_i + i$ is constant across the indices $i$ that satisfy $\rho_i \neq \rho_{i+1}$, letting $\rho_{|\rho|} = 0$.
\end{lemma}

\begin{theorem}
$|I_k(132, 213)$| counts Gorenstein partitions of $k$.  This is listed as \texttt{A117629} on the OEIS.
\end{theorem}

\begin{theorem}
Let $\mu \vDash s$ denote that $\mu$ is a composition of $s$. Then $|I_k(132, 213)|$ has generating function $\sum_{s \geq 0} \sum_{\mu \vDash s, |\mu| \neq 1} x^{\binom{s}{2} - \sum_{m \in \mu} \binom{m}{2}}$. This means $|I_k(132, 213)|$ also enumerates finite sequences of positive integers of length $> 1$ such that $k$ equals the second elementary symmetric function of the values of the sequence, as noted in the OEIS entry.
\end{theorem}

The generating function is not useful for actually computing new terms in the sequence, but we can use the following recurrence instead. By ignoring all but the first and last $\sqrt{n}$ summands in the recurrence below, as they are zero, we can compute the $n$-th value in $\mathcal{O}(n^{2.5})$ time and $\mathcal{O}(n^2)$ space.

\begin{theorem}
The number of Gorenstein partitions with sum $n$, and thus also the number of elements in $|I_n(132, 213)|$, is given by the sum $\sum_{d = 0}^n f(n, d)$ where
\[f(n, d) = \begin{cases}
0 \text{ if } n < 0 \\
1 \text{ if } n = 0  \\
\sum_{k = 1}^d f(n - k(d + 1 - k), d - k) \text{ otherwise }
\end{cases}\]
\end{theorem}

\begin{theorem}
$|I_k(132, 231)|$ counts partitions on $k$ elements with distinct parts, this is listed as \texttt{A000009} on the OEIS.
\end{theorem}

\begin{theorem}
$|I_k(132, 321)|$ counts partitions on $k$ elements with equal values. This is in turn equal to the number of divisors of $k$. This is listed as \texttt{A000005} on the OEIS.
\end{theorem}

\section{More than two patterns}

Most of the remaining pattern combinations are trivially deduced as some subset of the patterns forces the sequence to die out or contain only very specific permutations. We consider here the complement of those cases.

\begin{theorem}
$|I_k(123, 132, 231)| = 1$.
\end{theorem}

\begin{theorem}
$|I_k(123, 132, 213)|$ enumerates the Pascal triangle, read by diagonals, offset by two elements. This means it reads the binomials $\binom{n}{k}$ in increasing order by the sum $n + k$, with each set being read in increasing order by $n$, and $|I_0(123, 132, 213)|$ starts at $\binom{1}{1}$. Furthermore its generating function can be written as $1 + \sum_{d \geq 3} x^{\binom{d - 1}{2}} \sum_{n = 2}^d \binom{n}{d - n} x^{n - 2}$.
\end{theorem}

\begin{theorem}
$|I_k(132, 213, 231)|$ counts the odd divisors of $k$, which is listed as \texttt{A001227} on the OEIS.
\end{theorem}

\begin{theorem}
$|I_k(132, 213, 321)| = |I_k(132, 321)|$.
\end{theorem}

\begin{theorem}
$|I_k(123, 132, 213, 231)|$ enumerates the Pascal triangle, except all values $> 1$ are replaced by $0$. This is listed as on \texttt{A103451} on the OEIS and has the generating function $\sum_{i\geq 0} x^{i(i+1)/2} + x^{(i+1)(i+4)/2}$.
\end{theorem}

\section*{Acknowledgements}

The algorithm in \cite{invalg} was used to generate elements of all the sequences above, which helped tremendously in finding the formulas and other results in this paper.

\nocite{*}
\bibliographystyle{eptcs}
\bibliography{generic}

\begin{thebibliography}{1}
\providecommand{\bibitemdeclare}[2]{}
\providecommand{\surnamestart}{}
\providecommand{\surnameend}{}
\providecommand{\urlprefix}{Available at }
\providecommand{\url}[1]{\texttt{#1}}
\providecommand{\href}[2]{\texttt{#2}}
\providecommand{\urlalt}[2]{\href{#1}{#2}}
\providecommand{\doi}[1]{doi:\urlalt{https://doi.org/#1}{#1}}
\providecommand{\eprint}[1]{arXiv:\urlalt{https://arxiv.org/abs/#1}{#1}}
\providecommand{\bibinfo}[2]{#2}

\bibitemdeclare{misc}{fountain}
\bibitem{fountain}
\bibinfo{author}{Peter \surnamestart Bala\surnameend} (\bibinfo{year}{2019}):
  \emph{\bibinfo{title}{FOUNTAINS OF COINS AND SKEW FERRERS DIAGRAMS}}.
\newblock \urlprefix\url{https://oeis.org/A161492/a161492_1.pdf}.

\bibitemdeclare{article}{321}
\bibitem{321}
\bibinfo{author}{E.~\surnamestart Barcucci\surnameend}, \bibinfo{author}{A.Del
  \surnamestart Lungo\surnameend}, \bibinfo{author}{E.~\surnamestart
  Pergola\surnameend} \& \bibinfo{author}{R.~\surnamestart Pinzani\surnameend}
  (\bibinfo{year}{2001}): \emph{\bibinfo{title}{Some permutations with
  forbidden subsequences and their inversion number}}.
\newblock {\slshape \bibinfo{journal}{Discrete Mathematics}}
  \bibinfo{volume}{234}(\bibinfo{number}{1}), pp. \bibinfo{pages}{1--15},
  \doi{10.1016/S0012-365X(00)00359-9}.
\newblock
  \urlprefix\url{https://www.sciencedirect.com/science/article/pii/S0012365X00003599}.

\bibitemdeclare{article}{akc}
\bibitem{akc}
\bibinfo{author}{Anders \surnamestart Claesson\surnameend},
  \bibinfo{author}{Vít \surnamestart Jelínek\surnameend} \&
  \bibinfo{author}{Einar \surnamestart Steingrímsson\surnameend}
  (\bibinfo{year}{2012}): \emph{\bibinfo{title}{Upper bounds for the
  Stanley–Wilf limit of 1324 and other layered patterns}}.
\newblock {\slshape \bibinfo{journal}{Journal of Combinatorial Theory, Series
  A}} \bibinfo{volume}{119}(\bibinfo{number}{8}), pp.
  \bibinfo{pages}{1680--1691}, \doi{10.1016/j.jcta.2012.05.006}.
\newblock
  \urlprefix\url{https://www.sciencedirect.com/science/article/pii/S0097316512000891}.

\bibitemdeclare{article}{invalg}
\bibitem{invalg}
\bibinfo{author}{Scott \surnamestart Effler\surnameend} \&
  \bibinfo{author}{Frank \surnamestart Ruskey\surnameend}
  (\bibinfo{year}{2003}): \emph{\bibinfo{title}{A CAT algorithm for generating
  permutations with a fixed number of inversions}}.
\newblock {\slshape \bibinfo{journal}{Information Processing Letters}}
  \bibinfo{volume}{86}(\bibinfo{number}{2}), pp. \bibinfo{pages}{107--112},
  \doi{10.1016/S0020-0190(02)00481-7}.
\newblock
  \urlprefix\url{https://www.sciencedirect.com/science/article/pii/S0020019002004817}.

\bibitemdeclare{article}{cons}
\bibitem{cons}
\bibinfo{author}{Thomas~E. \surnamestart Mason\surnameend}
  (\bibinfo{year}{1912}): \emph{\bibinfo{title}{On the Representation of an
  Integer as the Sum of Consecutive Integers}}.
\newblock {\slshape \bibinfo{journal}{The American Mathematical Monthly}}
  \bibinfo{volume}{19}(\bibinfo{number}{3}), pp. \bibinfo{pages}{46--50},
  \doi{10.1080/00029890.1912.11997664}.

\bibitemdeclare{article}{seminal}
\bibitem{seminal}
\bibinfo{author}{Rodica \surnamestart Simion\surnameend} \&
  \bibinfo{author}{Frank~W. \surnamestart Schmidt\surnameend}
  (\bibinfo{year}{1985}): \emph{\bibinfo{title}{Restricted Permutations}}.
\newblock {\slshape \bibinfo{journal}{European Journal of Combinatorics}}
  \bibinfo{volume}{6}(\bibinfo{number}{4}), pp. \bibinfo{pages}{383--406},
  \doi{10.1016/S0195-6698(85)80052-4}.
\newblock
  \urlprefix\url{https://www.sciencedirect.com/science/article/pii/S0195669885800524}.

\bibitemdeclare{article}{goren}
\bibitem{goren}
\bibinfo{author}{Richard~P \surnamestart Stanley\surnameend}
  (\bibinfo{year}{1978}): \emph{\bibinfo{title}{Hilbert functions of graded
  algebras}}.
\newblock {\slshape \bibinfo{journal}{Advances in Mathematics}}
  \bibinfo{volume}{28}(\bibinfo{number}{1}), pp. \bibinfo{pages}{57--83},
  \doi{10.1016/0001-8708(78)90045-2}.
\newblock
  \urlprefix\url{https://www.sciencedirect.com/science/article/pii/0001870878900452}.

\end{thebibliography}
\end{document}